\documentclass[11pt,a4paper]{article}
\pdfoutput=1	
\usepackage{multirow}
\usepackage{multicol}
\usepackage{alpha}
\usepackage{macros}
\usepackage{graphicx}
\usepackage{amsmath}
\usepackage{amssymb}
\usepackage{sidecap}
\usepackage{wrapfig}
\usepackage{comment}
\usepackage{color}

\newcommand{\SF}{Schr\"odinger functional\ }

\newcommand{\nf}{N_{\rm f}}
\newcommand{\Vub}{V_{\rm ub}}
\newcommand{\mbeauty}{m_{\rm b}}
\newcommand{\mlight}{m_{\rm l}}
\newcommand{\Mbeauty}{M_{\rm b}}
\newcommand{\rmd }{\mathrm{d}}
\newcommand{\vecp}{{\bf p}}
\newcommand{\vecq}{{\bf q}}
\newcommand{\vecx}{{\bf x}}
\newcommand{\cC}{{\cal C}}
\providecommand{\Ct}{\ensuremath{\mathcal C^{\B\to\K} } }

\providecommand{\B}{\ensuremath{{\textnormal{B}_{\textnormal{s}}} }}
\providecommand{\K}{\ensuremath{\textnormal{K}} }
\providecommand{\RGI}{\ensuremath{\textnormal{stat,RGI}} }
\providecommand{\hqetb}{\ensuremath{\textnormal{stat,bare}} }
\DeclareMathOperator{\e}{e}
\DeclareMathOperator{\im}{i}

\newcommand{\rmO}{\mathrm{O}}

\def\bec{\begin{center}}
\def\eec{\end{center}}
\def\beq{\begin{equation}}
\def\eeq{\end{equation}}
\def\bes{\begin{eqnarray}}
\def\ees{\end{eqnarray}}

\newcommand{\simas}[1]{\raisebox{-.1ex}{
            $\stackrel{\small{#1}}{\sim}$}}

\begin{document}
\preprintno
{
DESY~16-009\\
DAMTP-2016-13\\[3em]
}

\title{Continuum limit of the leading-order HQET 
form factor in $\mathrm B_\mathrm s \to \mathrm K \ell \nu$ decays }

\collaboration{\includegraphics[width=2.8cm]{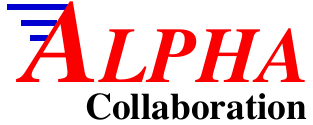}}

\author[desy]{Felix~Bahr}
\author[desy]{Debasish~Banerjee}
\author[desy,dresden]{Fabio~Bernardoni}
\author[cambridge]{Anosh~Joseph}
\author[desy]{Mateusz~Koren}
\author[desy]{Hubert~Simma}
\author[desy]{Rainer~Sommer}

\address[desy]{John von Neumann Institute for Computing (NIC), DESY, Platanenallee~6, D-15738~Zeuthen, Germany}
\address[dresden]{Medizinische Fakult\"at, Carl Gustav Carus, TU Dresden, Fetscherstra\ss e~74, D-01307~Dresden, Germany}
\address[cambridge]{Department of Applied Mathematics and Theoretical Physics (DAMTP),
University of Cambridge, Cambridge, CB3 0WA, UK
}
\begin{abstract}

We discuss the computation of form factors for 
semi-leptonic decays of $\rm B$-, $\B$- mesons in lattice QCD. 
Considering in particular the example of the static $\B$ form factors we demonstrate 
that after non-perturbative renormalization the 
continuum limit can be taken with confidence. The resulting
precision is of interest for extractions of $V_\mathrm{ub}$.
The size of the corrections of order $1/\mbeauty$ is just estimated 
at present but it is expected that their inclusion does 
not pose significant difficulties. 

\end{abstract}

\maketitle

\section{Introduction}
\label{s:intro}

Weak decays of B-mesons are a very important piece in the puzzle
of understanding
about how well the Standard Model of particle physics describes Nature. 
One relevant question concerns the determination of the Cabibbo-Kobayashi-Maskawa
matrix element $\Vub$ from different decays. This fundamental parameter of
the Standard Model is not known very precisely yet. Testing for consistent values 
provides a check of the Standard Model. In fact results extracted from
inclusive decays agree with those from different exclusive decays, like 
$\mathrm B \rightarrow \pi \ell \nu$ or $B\to\tau\nu$ \cite{Agashe:2014kda,Abdesselam:2014hkd,Adachi:2012mm}, 
only after stretching the presently estimated uncertainties by around a factor three. 
We avoid calling this a three-sigma tension since the uncertainties are largely
systematic, coming from the theoretical computation of form factors in
lattice QCD on one side and the perturbative treatment of inclusive decays on
the other side. But also experimental uncertainties contribute.

In this letter we consider the determinations of semi-leptonic form factors 
for \mbox{$\B$-mesons} 
from lattice QCD. A review with some discussion of the challenges involved is
found in \cite{Aoki:2013ldr}. It appears that the most relevant challenge is 
the presence of a (large) mass scale
$\mbeauty\sim 5\;\GeV$. 
Together with inverse lattice
spacings below $4\;\GeV$ this distorts the continuum physics considerably in a straight 
application of lattice QCD. We do not want to review here this issue in detail,
but just mention that this leads one to consider effective field theories for 
the b-quark or extrapolations in its mass, again guided
by effective field theory considerations. 
The most advanced computations
\cite{Lattice:2015tia
,Flynn:2015mha
,Bouchard:2014ypa
,Colquhoun:2015mfa
}, use either a relativistic heavy quark action or
employ non-relativistic QCD on the 
lattice. There the challenge is twofold. 
First, a fully non-perturbative renormalization program for the heavy-light currents
does not (yet) exist. It is replaced by ``mostly non-perturbative'' renormalization \cite{ElKhadra:1997hq,Harada:2001fi}, where the factor $Z_\mathrm{hl}/\sqrt{Z_\mathrm{hh}Z_\mathrm{ll}}$ 
is taken from 1-loop perturbation theory and this approximation is expected to
be a good one \cite{ElKhadra:1997hq,Harada:2001fi}; alternatively straight 1-loop perturbation theory is used.
Second, discretisation errors are estimated only by power-counting arguments because
continuum limit extrapolations may involve a complicated 
functional dependence on the lattice spacing. As a consequence we are not aware
of the computation of a non-perturbatively renormalized heavy-light form factor 
extrapolated to the continuum. 

In order to place the present work into context, let us briefly list the 
steps which are necessary to come to a trustworthy result of interest to phenomenology:
\begin{itemize}\setlength{\itemsep}{-1mm}
\item[a)] obtain the {\em ground state} matrix elements 
  $\langle \K | V^\mu(0) | \Bs \rangle$ that mediate the transition, 
\item[b)] renormalize the currents (and thus matrix elements)
    and, if an effective theory is used, relate them to QCD (``matching''),
\item[c)] take their continuum limit,
\item[d)] extrapolate to the quark masses realized in Nature,
\item[e)] map out the $q^2$ dependence.
\end{itemize}
Here we demonstrate our solutions to a)--c). We are very brief 
about our specific choices in a), even though the extraction of the
ground-state-to-ground-state matrix element which gives the 
leading form factor, $\fperp$,{}
is delicate since excited state contributions
have signifi\-cant amplitudes. Details on this more technical but important issue are relegated
to a companion paper \cite{semilept:preparation}.
Steps d)--e) will follow in due course.

We concentrate on the non-perturbative renormalization and the 
continuum limit albeit only in the leading order of Heavy Quark
Effective Theory (HQET).
We consider a single value of the momentum transfer and a single value
of the light dynamical quark masses with only two degenerate dynamical
flavors. These 
restrictions mean that our computation does not immediately advance phenomenology,
but since a continuum limit was not taken before we can see for the first time
how it works, and
the result provides a cross check on the uncertainty estimates of 
previous computations. We will find that 
with our discretisation and with non-perturbative renormalization, the continuum limit (for Kaon momentum around $0.5\;\GeV$) is smooth. Given that the inclusion
of $1/\mbeauty$ effects in the systematic treatment of HQET \cite{Sommer:2015hea}
was not a severe problem (apart from a lot of work) 
in other quantities \cite{Bernardoni:2013xba,Bernardoni:2014fva,Bernardoni:2015nqa} we are very encouraged to complete the started
programme towards phenomenologically relevant results. 
%

\section{Form Factors}
\label{s:ff}

We consider the decay $\mathrm B_\mathrm s \to \mathrm K \ell \nu$. 
Working at the leading order in the weak interactions, the
transition amplitude factorises into a straightforward leptonic 
amplitude and the QCD matrix element 
with the equivalent form-factor decompositions
\bes
  && \langle \K (p_\K)| V^\mu(0) | \Bs(p_{\Bs}) \rangle =
  \nonumber \\
  && \hspace*{3em} = 
  \left(p_\Bs+p_\K - \frac{m_\Bs^2-m_\K^2}{q^2} q\right)^\mu \cdot f_+(q^2)
  + \frac{m_\Bs^2-m_\K^2}{q^2} q^\mu \cdot f_0(q^2)
  \nonumber \\
  && \hspace*{3em} = 
  \sqrt{2m_{\Bs}} \Bigl[ v^\mu \cdot \fparallel(p_\K \cdot v) + p_\perp^\mu \cdot \fperp(p_\K \cdot v) \Bigr]
  \label{e:ffspp}\,.
\ees
The last line, with
\bes
  \nonumber 
 v^\mu = p_{\Bs}^\mu/ m_{\Bs}\,, \quad
  p_\perp^\mu  = p^\mu_\K  - (v \cdot p_\K)\, v^\mu \,,
\ees
defines the form factors, $\fparallel$ and $\fperp$.
The usual squared momentum transfer
$q^2 = (q^0)^2 - \vecq^2$, with $q^\mu \equiv p_{\Bs}^\mu - p_\K^\mu$,
is here replaced by 
\beq
   p_\K \cdot v = \frac{m_{\Bs}^2 + m_\K^2 - q^2}{2 m_{\Bs}}\,,
\eeq
which at fixed Kaon four-momentum $p_\K$ is independent of the mass of the b-quark. 
This property, together with 
the factor $(2 m_{\Bs})^{1/2}$ in \eqref{e:ffspp},
is convenient to discuss the 
behavior of the amplitude at large mass of the b-quark. 
It removes the mass-dependence
of the (standard) relativistic normalization 
$\langle \Bs(p') | \Bs(p) \rangle = 2E(\vecp) (2\pi)^3 \delta(\vecp-\vecp')$
of the state of the heavy meson. Since the current  
$V^\mu(x) \equiv \bar{\psi}_{\rm u}(x) \gamma^\mu \psi_{\rm b}(x)$
translates into heavy quark effective (mass-independent) fields 
with only a logarithmically mass-dependent conversion function,
the form factors $\fparallel$ and $\fperp$ 
scale only logarithmically with the mass in the limit of large b-quark mass. 

Choosing for the remainder of this letter 
the rest-frame of the $\Bs$-meson with $v^\mu=(1,0,0,0)$ 
as a reference frame,
the invariant kinematic variable is just 
$p_\K \cdot v=E_\K$, the energy of the final-state
pseudo-scalar.
Upon neglecting terms proportional to $m_\ell^2/m_\Bs^2$ and
$m_\ell^2/q^2$ ($m_\ell$ being the mass of the final-state lepton),
the differential decay rate is then given by 
\bes
  {\rmd \Gamma(\Bs\to \mathrm K \ell \nu) \over \rmd q^2} 
  &=& {G_\mathrm{F}^2 \over 24 \pi^3} |\Vub|^2
  |\vecp_\K|^3 [f_+(q^2)]^2 \,.
  \label{e:dGamma}
\ees
A comparison of \eqref{e:dGamma}
with the experimentally measured
rate allows for a determination of
$|\Vub|$ once the form factors are known. They need to be determined
at a single value (or ideally
in a range)
of $E_\K$ where overlap with experimental data exists.

In our frame ($\vecp_\B=0$), the form factors are obtained from
the (QCD) matrix elements
\bes
   (2m_{\Bs})^{-1/2} \langle \K (p_\K)| V^0(0) | \Bs \rangle &=& 
 \fparallel(E_\K) \,, \\
   (2m_{\Bs})^{-1/2} \langle \K (p_\K)| V^k(0) | \Bs \rangle &=& 
 p_\K^k \fperp(E_\K)  \,.
\ees
With the above normalization, \eq{ffspp}, they have an HQET expansion 
\bes
 \fparallel(E_\K) &=& C_\mathrm{V_0}(\Mbeauty/\Lambda_\msbar)  
 \fparallel^\RGI(E_\K)\cdot [1 + \rmO(1/\mbeauty)]\,,
 \label{e:fparhqet}
 \\
\fperp(E_\K) &=& C_\mathrm{V_k}(\Mbeauty/\Lambda_\msbar)  \,
 \fperp^\RGI(E_\K)\cdot [1 + \rmO(1/\mbeauty)]
 \label{e:fperhqet}
\ees
without factors that involve a power of the quark mass. Rather 
the r.h.s.\ depend logarithmically  on 
the mass of the heavy quark, due to
the matching of HQET to QCD. 
In our notation, 
\bes
  V_{0,k}^\RGI = Z^\RGI_{V_{0,k}} V_{0,k}^\mathrm{stat}
   \label{e:phirgi}
\ees
are the renormalization group invariant (RGI) operators in HQET, and
the conversion functions $C_x$ connect (the matrix elements of) 
$V_0^\RGI$ and $V_k^\RGI$ to the ones in QCD, see \cite{Heitger:2003nj,Sommer:2010ic}.

The functions $C_x$ are known with 2-loop matching 
(for short ``2-loop''), i.e.\  
up to $\alpha(\mbeauty)^3$ 
corrections in continuum perturbation theory
\cite{Shifman:1987sm,Politzer:1988wp,BroadhGrozin,ChetGrozin,Ji:1991pr,BroadhGrozin2,Gimenez:1992bf,Bekavac:2009zc}.
We use them here with the RGI b-quark mass $\Mbeauty$ and the
$\Lambda$-parameter determined in the theory with two dynamical
flavors \cite{Bernardoni:2013xba,Fritzsch:2012wq}, i.e. 
$\Mbeauty/\Lambda_\msbar = 21.2(1.2)$, where the uncertainty of 
$\Lambda$ dominates. The conversion functions then evaluate to\footnote{They are conveniently summarized in \cite{Fritzsch:2015eka}.} 
\bes
   C_\mathrm{V_0}(\Mbeauty/\Lambda_\msbar)  = 1.214(6)(13)\,,\label{e:Cv0}
   \\
   C_\mathrm{V_k}(\Mbeauty/\Lambda_\msbar) = 1.134(7)(47)\,,\label{e:Cvk}
\ees
where the second quoted uncertainty is estimated as the 
difference between 2-loop and 1-loop.
It is not entirely clear whether this is a 
conservative estimate of the perturbative error (see sect.~2.3 of \cite{Sommer:2015hea}). 

Let us clarify the difference to standard 1-loop renormalization of heavy-light form factors. 
We renormalize the HQET currents in \eqref{e:phirgi}  non-perturbatively,
thus the conti\-nuum limit of their matrix elements is not affected by any perturbative uncertainty. 
Then, in the continuum, the factor $C_x$ is known only perturbatively, but
to one more power of 
$\alpha_\mathrm{s}$ than what is available for the total 
$Z_x=C_x\times Z^\RGI_x$
in other approaches. Thus, even if we quote an uncertainty of up to five percent for
the renormalization, this is an $\rmO(\alpha^3)$ uncertainty where usually
it is $\rmO(\alpha^2)$. 
In the future the  ALPHA collaboration will non-perturbatively match HQET and QCD \cite{Blossier:2012qu}
also for the vector currents \cite{DellaMorte:2013ega}. One then obtains directly $Z_x=C_x\times Z^\RGI_x$
with full non-perturbative precision.

We now proceed to the numerical evaluation of the $\mbeauty$-independent 
RGI matrix elements $h_x^\RGI$, which are not affected
by perturbative errors or ambiguities.

\section{Lattice Calculation}
\label{s:latt}

\subsection{Framework and Renormalization}
\label{s:ren}

For our first numerical investigation of the problem, we choose $\nf=2$ 
flavors of quarks. The prime reason for this choice is that in
a related project, the non-perturbative matching of HQET to QCD 
for the currents $V_0,V_k$ is being carried out at the order
$1/\mbeauty$ \cite{DellaMorte:2013ega,Hesse:2012hb,Korcyl:2013ara,Korcyl:2013ega}. 
Once this is complete, we will be able to include
the $1/\mbeauty$ corrections with little additional effort. 
For now, 
we remain at the lowest order of HQET, namely the static order.

The b-quark is then replaced by a static quark \cite{Eichten:1989zv}
labelled ``h''. Two different discretisations, HYP1 and HYP2,
are chosen \cite{DellaMorte:2005yc}. These have moderate discretisation errors and
a much improved signal-to-noise ratio compared to the classic Eichten-Hill static
quark action\cite{Eichten:1989zv}. 
The bare currents 
\begin{subequations}
\begin{align}
 V_0^\textnormal{stat} &=  \overline \psi_\textnormal u \gamma_0          \psi_\textnormal h + ac_{\textnormal V_0}(g_0)
 \overline \psi _\textnormal l \sum_l \overleftarrow\nabla^\textnormal S_l \gamma_l          \psi_\textnormal h, \\
V_k^\textnormal{stat} &=  \overline \psi_\textnormal u \gamma_k          \psi_\textnormal h - ac_{\textnormal V_k}(g_0) 
\overline \psi _\textnormal l \sum_l \overleftarrow\nabla^\textnormal S_l \gamma_l \gamma_k \psi_\textnormal h,
\end{align}
\label{deltav}
\end{subequations}
are form-identical to the ones in QCD, apart from the
$\rmO(a)$ improvement terms 
($\overleftarrow\nabla^\textnormal S$ denotes the symmetric covariant derivative acting on the field to the left).
The coefficients\footnote{Spin symmetry leads to the identity $c_{\textnormal V_k} = c_{\textnormal A_0}$. 
Ref. \cite{Grimbach:2008uy} uses the notation $c_{\textnormal A}^\mathrm{stat},\;c_{\textnormal V}^\mathrm{stat}$ for
$c_{\textnormal A_0},\;c_{\textnormal V_0}$.}, 
$c_{\textnormal V_0}$, $c_{\textnormal V_k}$, are known
to 1-loop order: $c_x=c_x^{(1)} g_0^2 + \rmO(g_0^4)$ with \cite{Grimbach:2008uy}
\begin{subequations}
\begin{align}
 \text{HYP1: }\; c_{\textnormal V_k}^{(1)} = 0.0029(2) \,,\quad
                 c_{\textnormal V_0}^{(1)} = 0.0223(6) \ \\
 \text{HYP2: }\; c_{\textnormal V_k}^{(1)} = 0.0518(2) \,,\quad
                 c_{\textnormal V_0}^{(1)} = 0.0380(6) \,. 
\end{align}
\end{subequations}
The multiplicative renormalization of the currents \eq{phirgi} can be written as
\begin{align}
 V_0^\mathrm{\RGI} &= Z_{A_0}^\RGI (g_0) 
   Z_\mathrm{V/A}^\mathrm{stat} (g_0) V_0^\mathrm{stat}, 
   \label{v0renorm} \\
 V_k^\mathrm{\RGI} &= Z_{A_0}^\RGI (g_0)  
   V_k^\mathrm{stat}\,, 
   \label{vkrenorm}
\end{align}
where we have made explicit that 
$ Z^\RGI_{V_{k}}= Z_{A_0}^\RGI$ because
$V_k^\mathrm{stat}$ renormalizes exactly as $A_0^\mathrm{stat}$ due to the spin symmetry of HQET, while for $V_0$ an extra
factor $Z_\mathrm{V/A}^\mathrm{stat} (g_0)$  originates from
the broken chiral symmetry of Wilson fermions.
We use the non-perturbative results 
\bes
Z_{A_0}^\RGI (g_0) = R(\mu)\, 
  Z_\mathrm{A}^\mathrm{stat}(g_0,a\mu)
\ees
with \cite{DellaMorte:2006sv}
\bes
   R(\mu_0) = 0.880(7) \label{e:RGIfactor}
\ees
relating $A_0^\stat$, renormalized at scale $\mu_0=1/L_\mathrm{max}$ in the \SF scheme, to the RGI operator.
The factor $R$ is known non-perturbatively by running to very high
$\mu$ and continuum extrapolation.
In \tab{t:zastat}, the remaining piece
$Z_\mathrm{A}^\mathrm{stat}(g_0,a\mu_0)$ is reproduced from \cite{DellaMorte:2006sv}.
For the finite renormalization $Z_\mathrm{V/A}^\mathrm{stat}$
we use a range
\bes
  [Z_\mathrm{V/A}^\mathrm{stat} (g_0)]^{-1} = 0.97(3)\,.
  \label{e:zva}
\ees
This range is generous, because $Z_\mathrm{V/A}^\mathrm{stat}$ has been seen to be very close to one 
in the quenched approximation \cite{Palombi:2007dt}, and at 1-loop order, there is no $\nf$--dependence. 
Despite these arguments our range in \eqref{e:zva} 
is no more than an educated guess. 
This is adequate since $Z_\mathrm{V/A}^\mathrm{stat}$ affects only one of the $1/\mbeauty$ suppressed terms, which we just use as an illustration of the 
associated uncertainty. Again, we note that such issues 
will be eliminated when the non-perturbative matching is carried out.

\begin{table}
\begin{center}
\begin{tabular}{ccccc}
        & \multicolumn{2}{c}{$c_x=0$} & \multicolumn{2}{c}{$c_x=c_x^{(1)} g_0^2$} \\
$\beta$ &  HYP1  &  HYP2  & HYP1  &  HYP2 \\ \hline
$5.2$   & $0.7104(\hphantom{2}5)$ & $0.7920(\hphantom{2}5)$ & $0.7007(\hphantom{2}5)$ & $0.7432(\hphantom{2}5)$ \\
$5.3$   & $0.7057(27)$            & $0.7839(26)$            & $0.6965(27)$            & $0.7376(25)$ \\
$5.5$   & $0.6901(27)$            & $0.7597(26)$            & $0.6820(26)$            & $0.7218(24)$
\end{tabular}
\caption{Values for $Z_\mathrm{A}^\mathrm{stat}(g_0,a\mu_0) \times 0.880$. At
$\beta=5.2$ they are taken directly from table 4 of reference \cite{DellaMorte:2006sv}, at $\beta=5.3$ they are obtained using the interpolating polynomial of equation (B.3) and table 9 of that reference and for the $\beta=5.5$ 
numbers we performed a linear extrapolation of the ones at $\beta=5.29$ and $\beta=5.4$.} \label{t:zastat}
\end{center}
\end{table}

\subsection{Simulation Parameters}
\begin{table}
\begin{center}
\begin{tabular}{cccccccc}
id & $\beta$ & $L/a$ & $a$ [fm] &  $m_\pi$ [{\rm MeV}] & $N_{\rm cfg}$ & $\kappa_s$ & $\theta/(2\pi)$ \\ 
\hline
A5  & $5.2$  &  $32$ &  $0.0749(8)$  &$330$ & $1000$ & $0.13535$  &  $0.034$
\\
F6  & $5.3$ &  $48$ & $0.0652(6)$  &$310$  & $300$ & $0.13579$  &  $0.350$
\\
N6  & $5.5$ & $48$ &  $0.0483(4)$  &$340$  & $300$& $0.13631$  & $0$
\end{tabular}
\caption{Overview of the subset of $N_{\rm f} = 2 $ CLS ensembles on which we performed our measurements.  
Lattice spacings are taken from \cite{Lottini:2014zha}, an update of
\cite{Fritzsch:2012wq}.  All ensembles have $T=2L$ and 
$m_\pi L \geq 4$ where $m_\pi$ is the pion mass. $N_{\rm cfg}$ denotes the number of configurations on which we performed measurements.  The 
hopping parameter of the strange quark is denoted by $\kappa_s$.
The angle $\theta$ appears in the flavor-twisted boundary conditions, as explained in the text.} \label{tab:ens}
\end{center}
\end{table}

We base our investigation on a subset of the lattice gauge field configurations with
two degenerate flavors of improved Wilson fermions and Wilson gauge action generated by the Coordinated Lattice Simulations (CLS) effort
\cite{Fritzsch:2012wq}. 
The observables are computed on three ensembles, 
namely A5, F6 and N6, chosen to have roughly the same pion mass
but three different lattice spacings, see
\tab{tab:ens}. 
The quoted lattice spacings were determined 
from the Kaon decay constant $f_\K$ in reference \cite{Fritzsch:2012wq} and updated in reference \cite{Lottini:2014zha}.

Our choice of the gauge field ensembles fixes the 
degenerate up and down quark masses, $\mlight$. For the spectator strange (valence) quark mass we are free, however, to choose any
smooth function $\mstrange(\mlight)$ which passes through
the physical point. As in \cite{Fritzsch:2012wq} we define this 
function by fixing the
Kaon mass in units of the Kaon decay constant to its physical value
at our (and any) value of $\mlight$. We expect that this will
lead to a flat extrapolation to the physical value of $\mlight$, the ``physical point''.

We choose  $|\vecp_\K|=0.535\;\GeV$ which
corresponds to the minimum available momentum 
with periodic boundary conditions for all fields on the N6 lattice.  
On the other lattices we keep $\vecp_\K = (1,0,0)\,(2\pi +\theta)/L $ 
fixed by introducing flavor-twisted boundary conditions \cite{Bedaque:2004kc} 
$\psi_\mathrm{s}(x+L\hat{1}) = \rme^{i\theta}\psi_\mathrm{s}(x)$,
for the strange quark.  
The $\Bs$-meson is kept at rest by 
$\psi_\mathrm{h}(x+L\hat{1}) = \rme^{i\theta}\psi_\mathrm{h}(x)$,
and we remain with periodic boundary conditions for all other fields.
The numerical values for $\theta$ are listed in \tab{tab:ens}.
Our choice of $\vecp_\K$ yields a central value
of $q^2=21.22\;\GeV^2$ at all lattice spacings and an
error coming from the lattice spacing of $0.03\!-\!0.05\;\GeV^2$. 
Note that the flavor-twist is introduced only for  
quenched quarks.

\subsection{Correlation Functions and Matrix Elements}
\begin{figure}
\begin{center}
\scalebox{0.58}{\includegraphics{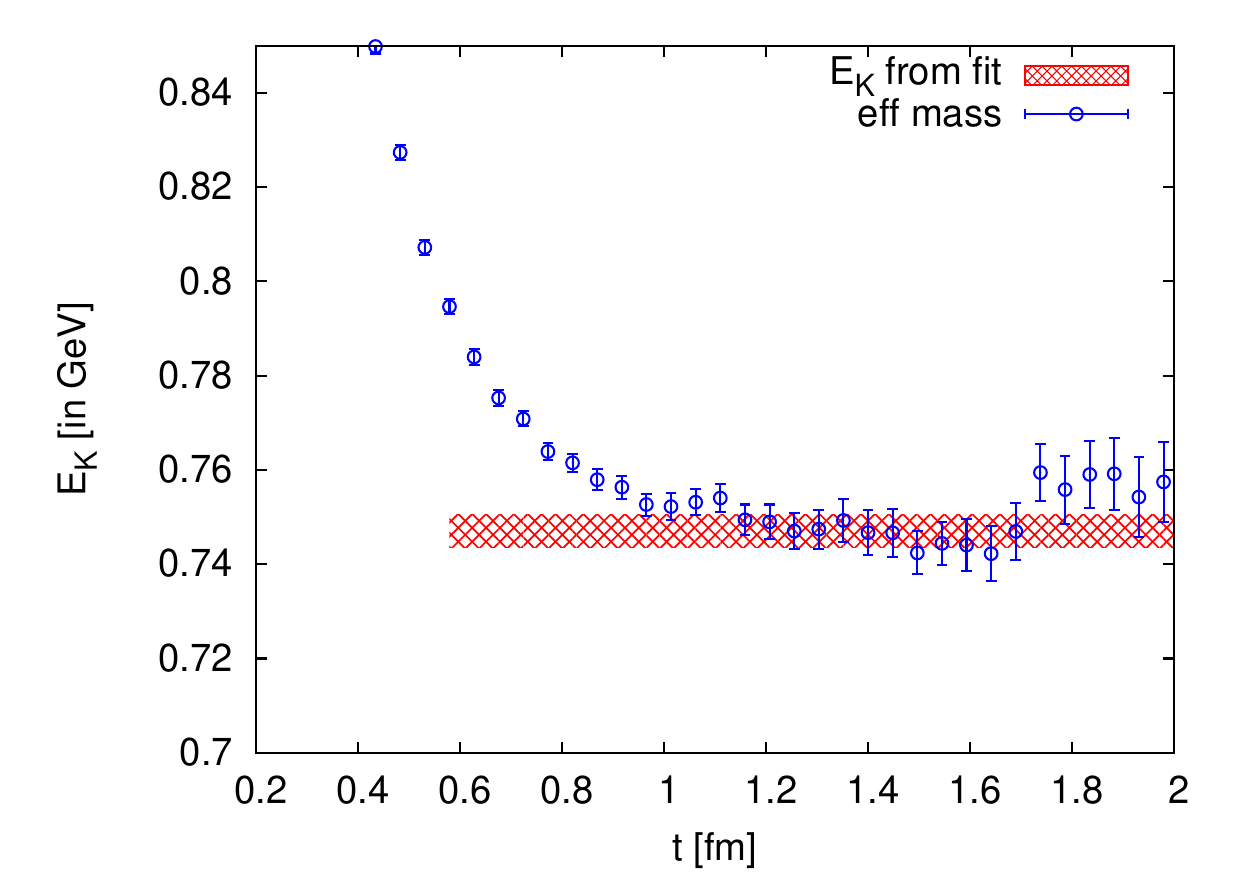} }
\scalebox{0.58}{\includegraphics{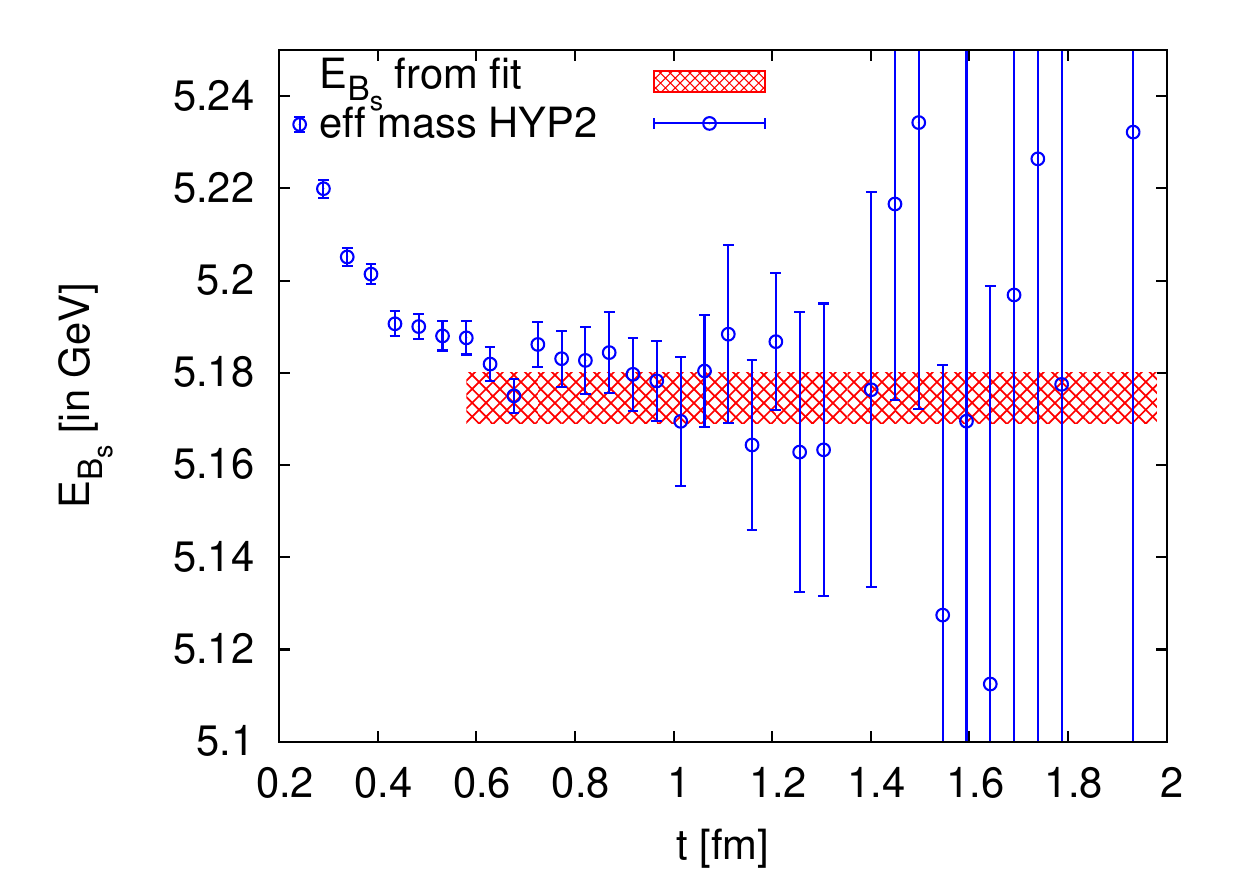} }

\caption{Effective energy of $\cC_\K$ (left) and $\cC_\B$ (right) on
  ensemble N6. Note that both panels have equal ranges on the
  corresponding axes. One can identify reasonable plateaus with small
  errors even though the Kaon carries a non-vanishing momentum and we
  have a static $\B$-meson.  The value for the ground state energy
  as obtained from a two-exponential fit is shown as a red
  band. Uncertainties shown here are only those of
  $E^\textnormal{stat}$ and $E_\K$ in lattice units, not the ones of
  $\mhbare$ and the lattice spacing. The data points are shown for the
  case of maximum smearing, while the fit involves all the smearing
  levels.}
\label{f:plat}
\end{center}
\end{figure}

We work with all-to-all light quark propagators 
\cite{Sommer:1994gg,Foster:1998vw}
implemented by a random U(1) source placed on each time slice 
(``full time dilution''\cite{Foley:2005ac}). 
While this is numerically costly, it significantly reduces the large-time variance. 
Together with the deflated solver 
\cite{ddhmc:webpage,Luscher:2007se,Frommer:2013fsa} that we use, it is thus very 
cost effective. Most notably, this feature of our computation 
provides access to all time separations of two-point and 
three-point functions on our lattices. For details we refer to
\cite{semilept:preparation}.

The two-point functions are defined as
\begin{align}
 \mathcal C^\K (y_0-x_0) & =  \sum\limits_{\vecy, \vecx} \e^{-\im \vec p \cdot (\vecy - \vecx)} \langle P_{\textnormal{s} \textnormal{u}} (y)P_{\textnormal{u} \textnormal{s}}(x)  \rangle ,  \label{ck} \\
  \mathcal C^\B_{ij} (y_0-x_0)  &=  \sum\limits_{\vecy, \vecx} \langle P^{(i)}_{\textnormal{s}\textnormal{b}}(y)P^{(j)}_{\textnormal{b}\textnormal{s}}(x)  \rangle , \label{cb} 
\end{align}
with the pseudoscalar density $P^{(i)}_{\textnormal{q}_1 \textnormal{q}_2} (x) = \overline{\psi}_{\textnormal{q}_1} (x) \gamma_5 \psi_{\textnormal{q}_2} (x) $. 
Indices $i$ and $j$ denote the use of different Gaussian wave functions for the light quarks
with a set of smearing parameters as in \cite{Bernardoni:2013xba}.
Up to terms which are exponentially suppressed in the time extent 
$T$ of the torus, we can 
parameterise $\mathcal C ^\K , \mathcal C^\B $ as
\begin{align}
 \mathcal C^\K (t_\K) &\simas{t_\K \gg a} \sum_n \big(\kappa^{(n)}\big)^2   \e^{-E_\K^{(n)} t_\K} \approx \big(\kappa^{(0)}\big)^2 \e^{-E^{(0)}_\K t_\K}, \label{e:ckek} \\
 \mathcal C^\B_{ij} (t_\B) &\simas{t_\B \gg a} \sum_n \beta_i^{(n)}  \beta_j^{(n)} \e^{-E_\B^{(n)} t_\B} \approx \sum\limits_{n=0}^N \beta_i^{(n)}  \beta_j^{(n)} \e^{-E_\B^{(n)} t_\B} ,  \label{e:cbeb} 
\end{align}
where we have denoted amplitudes by 
$\kappa^{(n)}=L^{3/2} (2E_\K)^{-1/2}\langle 0 | P_{\textnormal{u}\textnormal{s}}(0) |\K,n\rangle$ and
$\beta^{(n)}_i=L^{3/2} (2E_\B)^{-1/2}\langle 0 | P^{(i)}_{\textnormal{s}\textnormal{b}}(0) |\B,n\rangle$, while energies are labelled $E_\K^{(n)}, E_\B^{(n)}$.  The restrictions $t_x\gg a$ are required 
because we use an improved action where the positivity of the 
transfer matrix is not guaranteed. 
In a fit to these correlation functions, we use $t$-ranges denoted by $\tfit{K2}{min} \le t_\K \le \tfit{K2}{max}$
and $\tfit{B2}{min} \le t_\B \le \tfit{B2}{max}$, respectively, and
$N$ is the number of excited $\B$-meson states which we include.
Note that we have restricted ourselves here to only the ground state of the Kaon. 
There is a single, fixed, 
smearing level for the Kaon. In the Kaon two-point function,
ground-state dominance sets in at around 1.2~fm, while for the \B-meson 
this happens a bit earlier for our optimal (widest) Gaussian wave function. An illustration is found in \Fig{plat},
which also shows that a reasonably good precision is reached; plateaus are also clearly 
visible at the larger lattice spacings.

The three-point function 
\begin{equation}
 \mathcal C^{\B\to\K}_{\mu,j} (x^0_f-x^0_v, x^0_v-x^0_i) = \sum\limits_{\vecx_f, \vecx_v, \vecx_i} \e^{-\im \vecp \cdot (\vecx_f - \vecx _v)} \langle P_{\textnormal{s}\textnormal{u}}(x_f) V^\mu(x_v) P_{\textnormal{b}\textnormal{s}}^{(j)} (x_i) \rangle \label{c3}
\end{equation}
has a representation
\begin{align}
  \Ct_{\mu, i}(t_\K,t_\B) &\simas{t_\K,t_\B \gg a}  \sum\limits_m  \sum\limits_n \kappa^{(m)}  \varphi_\mu^{(m,n)} \beta_i^{(n)} \e^{-E^{(m)}_\K t_\K}  \e^{-E^{(n)}_\B t_\B}  \nonumber \\ 
& \approx \sum\limits_{n=0}^N \kappa^{(0)}  \varphi_\mu^{(0,n)} \beta_i^{(n)}   \e^{-E^{(0)}_\K t_\K} \e^{-E^{(n)}_\B t_\B}\,.
  \label{e:c3eb}
\end{align}
We perform a simultaneous fit to ${\mathcal C}^\K$, ${\mathcal C}^\B_{ij}$, and $\Ct_{\mu, i}$.
For the latter we consider $(t_\K,t_\B)$-values restricted to the rectangle with
$\tfit{K3}{min} \le t_\K \le \tfit{K3}{max,\mu }$ and $\tfit{B3}{min} \le t_\B \le \tfit{B3}{max,\mu}$.

The desired form factors are given by the ground-state matrix elements 
\begin{subequations}
\begin{align}
\fparallel^\hqetb  &=  \varphi^{(0,0)}_0 \sqrt{2E_\K}\, , \label{fparphi}  
\\
\fperp^\hqetb      &=  \frac{\varphi^{(0,0)}_k}{p_\K^k}\, \sqrt{2E_\K}\,.\label{fperphi}
\end{align}
\label{symffphi}
\end{subequations}
Their extraction from the data is rather delicate because statistical
errors grow with time separations. Due to our all-to-all computation
and the use of HYP1, HYP2 discretisations, we still have a precision
of better than two percent for $t_\K\lesssim2$~fm and
$t_\B\lesssim1.5$~fm in the two-point functions; for the three-point
function it drops below the two-percent level at around $t_\K + t_\B
\approx2$~fm (in the fitting region).
However, we find that $N=2$ excited states are necessary in our
fits to obtain a good description of the data and, in particular, a safe
extraction of the most important form factor $\varphi^{(0,0)}_k$.

\begin{figure}
\begin{center}
\vspace{-5pt}
\scalebox{0.29}{\includegraphics[angle=-90]{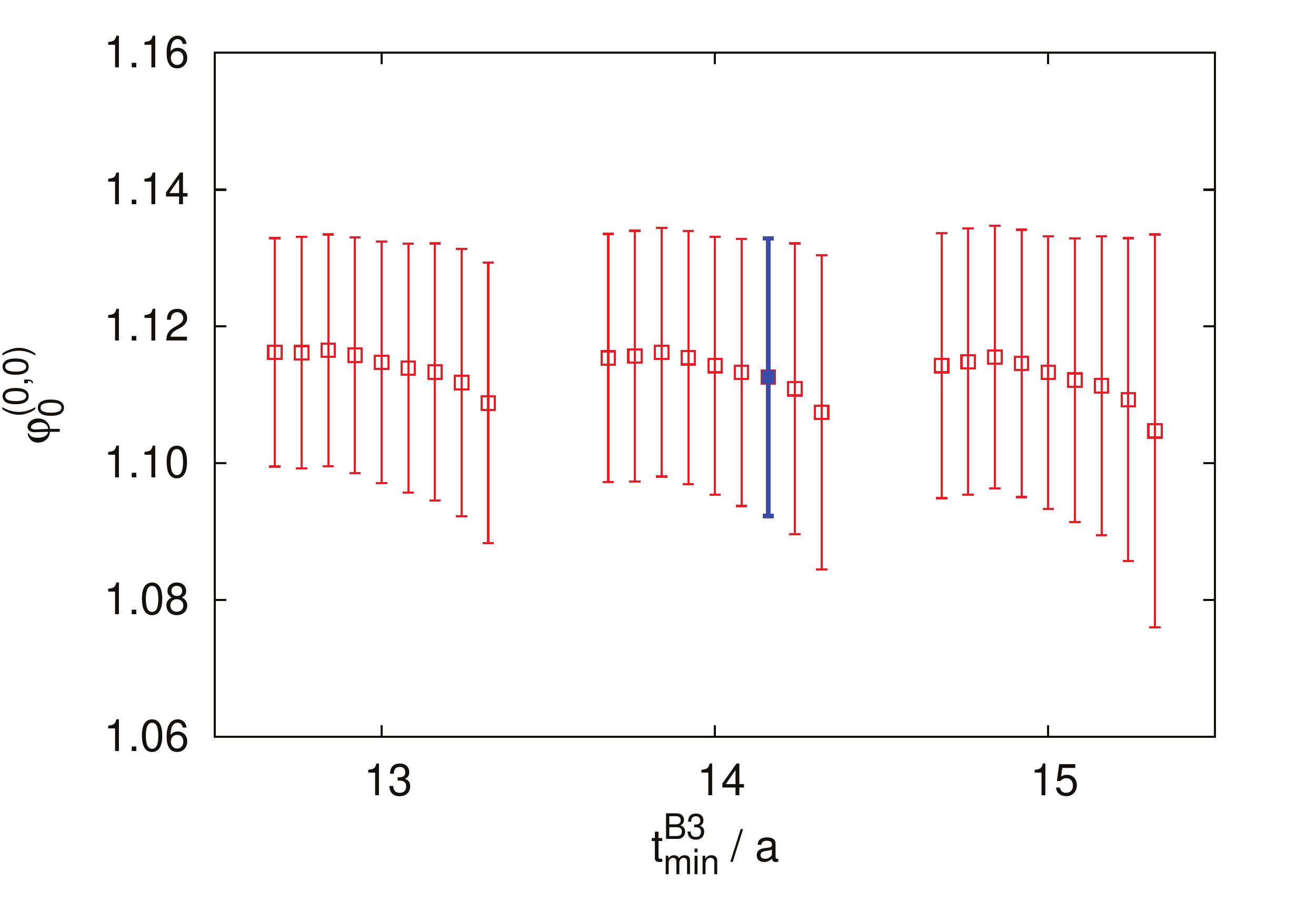} }
\scalebox{0.29}{\includegraphics[angle=-90]{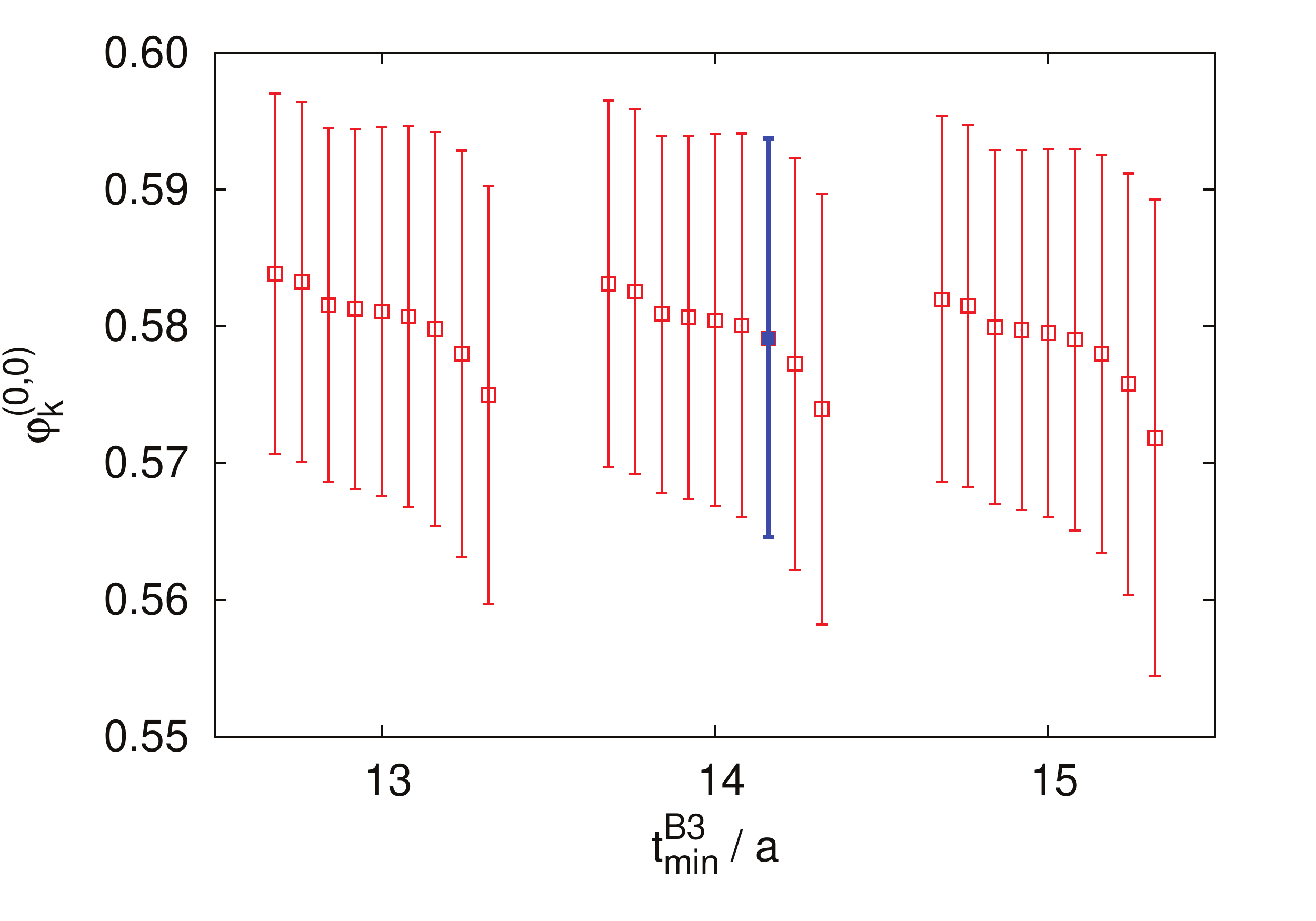} }
\vspace{-3pt}
\caption{
  Stability of the fit parameters $\varphi^{(0,0)}_0$ (left) and $\varphi^{(0,0)}_k$ (right) 
  on ensemble N6 with respect to variations of $\tfit{B3}{min}/a$ (different groups) 
  and of $\tfit{K3}{min}/a=11 \ldots 19$ (within the groups). In each panel, the value
  used to determine the bare form factor is marked with a filled square.}
\label{f:scatter}
\end{center}
\vspace{-6pt}
\end{figure}

To determine the reliability of the fits, we vary the boundaries
$\tfit{B2}{min}$, $\tfit{K3}{min}$, and $\tfit{B3}{min}$ of the fit
ranges and verify that the fit results remain unchanged within
errors. As an example, \fig{scatter} shows the dependence of the fit
results for $\varphi^{(0,0)}_0$ and $\varphi^{(0,0)}_k$ on
$\tfit{K3}{min}$ and $\tfit{B3}{min}$ (keeping $\tfit{B3}{min}-\tfit{B2}{min}=5a$ fixed). The
other boundaries are chosen to suppress the
effects of excited Kaon states ($\tfit{K2}{min}$), of noise
($\tfit{B2}{max}$), and of the finite time extent $T$. The latter two 
criteria considerably constrain our choice of $\tfit{K3}{max,\mu}$ and
$\tfit{B3}{max,\mu}$. 

\tab{t:fitrange} lists the fit ranges which we used for the HYP2 data. 
The bare ground-state matrix elements are shown in \tab{t:bareff}.
Further details will be described in \cite{semilept:preparation}.

\begin{table}[p]
\begin{center}
\begin{tabular}{lcccccccccc}
$\beta$ & $\tfit{K2}{min}$ & $\tfit{K2}{max}$ & $\tfit{B2}{min}$ & $\tfit{B2}{max}$  
& $\tfit{K3}{min}$ & $\tfit{K3}{max,0}$ & $\tfit{K3}{max,1}$ & $\tfit{B3}{min}$ & $\tfit{B3}{max,0}$ & $\tfit{B3}{max,1}$ 
\\
\hline
5.2 & 1.27 & 2.32 & 0.45 & 2.55& 0.82 & 1.35 & 1.05 & 0.67 & 1.57 & 1.35\\
5.3 & 1.43 & 3.06 & 0.46 & 2.54& 0.84 & 1.82 & 1.82 & 0.65 & 1.62 & 1.50\\
5.5 & 1.34 & 2.26 & 0.43 & 2.11& 0.82 & 1.34 & 1.06 & 0.67 & 1.34 & 1.44\\

\end{tabular}
\caption{Ranges of $t_\K$ and $t_\B$ (in fm) as used in our fits to HYP2 data.}
\label{t:fitrange}
\end{center}
\end{table}

\begin{table}
\begin{center}
\begin{tabular}{llllll}
 && \multicolumn{2}{c}{$c_x=0$} & \multicolumn{2}{c}{$c_x=c_x^{(1)}g_0^2$} \\
     &  $\beta$          & HYP1 &  HYP2  & HYP1 &  HYP2  \\
\hline
\\ $\fparallel^\hqetb$ [GeV$^{1/2}$] 
  \\   & 5.2   & $         1.38(3)$  & $         1.34(3)$  & $         1.38(3)$  & $         1.34(3)$ 
  \\   & 5.3   & $         1.42(4)$  & $         1.35(3)$  & $         1.42(4)$  & $         1.35(3)$ 
  \\   & 5.5   & $         1.40(3)$  & $         1.34(2)$  & $         1.40(3)$  & $         1.34(2)$ 
 \\ $\fperp^\hqetb$   [GeV$^{-1/2}$] 
  \\   & 5.2   & $         1.46(6)$  & $         1.33(6)$  & $         1.47(7)$  & $         1.39(6)$ 
  \\   & 5.3   & $         1.48(9)$  & $         1.35(8)$  & $         1.50(9)$  & $         1.40(8)$ 
  \\   & 5.5   & $         1.37(4)$  & $         1.25(3)$  & $         1.39(4)$  & $         1.31(3)$ 

\\
\end{tabular}
\caption{Unrenormalized form factors $\fparallel^\hqetb,\fperp^\hqetb$ for the specified 
discretisations.}
\label{t:bareff}
\end{center}
\end{table}

\begin{table}
\begin{center}
\begin{tabular}{llllll}
    & & \multicolumn{2}{c}{$c_x=0$} & \multicolumn{2}{c}{$c_x=c_x^{(1)}g_0^2$} \\
    &  $\beta$          & HYP1 &  HYP2  & HYP1 &  HYP2  \\
\hline
\\[-1mm] $\fparallel^\RGI$ [GeV$^{1/2}$] \\
     & 5.2   & $         1.01(4)$  & $         1.10(4)$  & $         1.00(4)$  & $         1.03(4)$ 
  \\   & 5.3   & $         1.03(5)$  & $         1.09(4)$  & $         1.02(4)$  & $         1.03(4)$ 
  \\   & 5.5   & $         0.99(4)$  & $         1.05(4)$  & $         0.98(4)$  & $         1.00(4)$ 
  \\   \multicolumn{2}{r}{continuum}   & $         0.98(5)$  & $         1.02(5)$  & $         0.97(5)$  & $         0.98(5)$ 
\\[2mm] $\fperp^\RGI$  [GeV$^{-1/2}$] \\
     & 5.2   & $         1.04(5)$  & $         1.05(5)$  & $         1.03(5)$  & $         1.04(5)$ 
  \\   & 5.3   & $         1.05(7)$  & $         1.06(6)$  & $         1.04(6)$  & $         1.04(6)$ 
  \\   & 5.5   & $         0.95(3)$  & $         0.95(3)$  & $         0.95(3)$  & $         0.94(3)$ 
  \\   \multicolumn{2}{r}{continuum}   & $         0.88(5)$  & $         0.88(5)$  & $         0.88(5)$  & $         0.87(5)$ 
  \\[2mm]   \multicolumn{2}{l}{$\fplusnum{1}$}   & $         1.78(8)$  & $         1.79(8)$  & $         1.78(8)$  & $         1.76(8)$ 

\\
\end{tabular}
\caption{Renormalized form factors $\fparallel^\RGI,\fperp^\RGI$ and their
continuum limits. The last line gives the conventional combination of
form factors, 
$f_+$.}
\label{t:rgiff}
\end{center}
\end{table}

\subsection{Continuum Limits}

The bare form factors, renormalized as explained in \sect{ren}, yield the RGI form factors listed in \tab{t:rgiff}. Their errors take account of 
all statistical correlations and autocorrelations as described in 
\cite{Bernardoni:2013xba} based on
\cite{Wolff:2003sm,Schaefer:2010hu}. 
For these non-perturbatively renormalized form factors (at fixed squared momentum 
transfer $q^2$, or Kaon energy $p_\K\cdot v$) the continuum limit can now be taken.

\begin{figure}
\begin{center}
\hspace{0cm}\scalebox{0.57}{\includegraphics{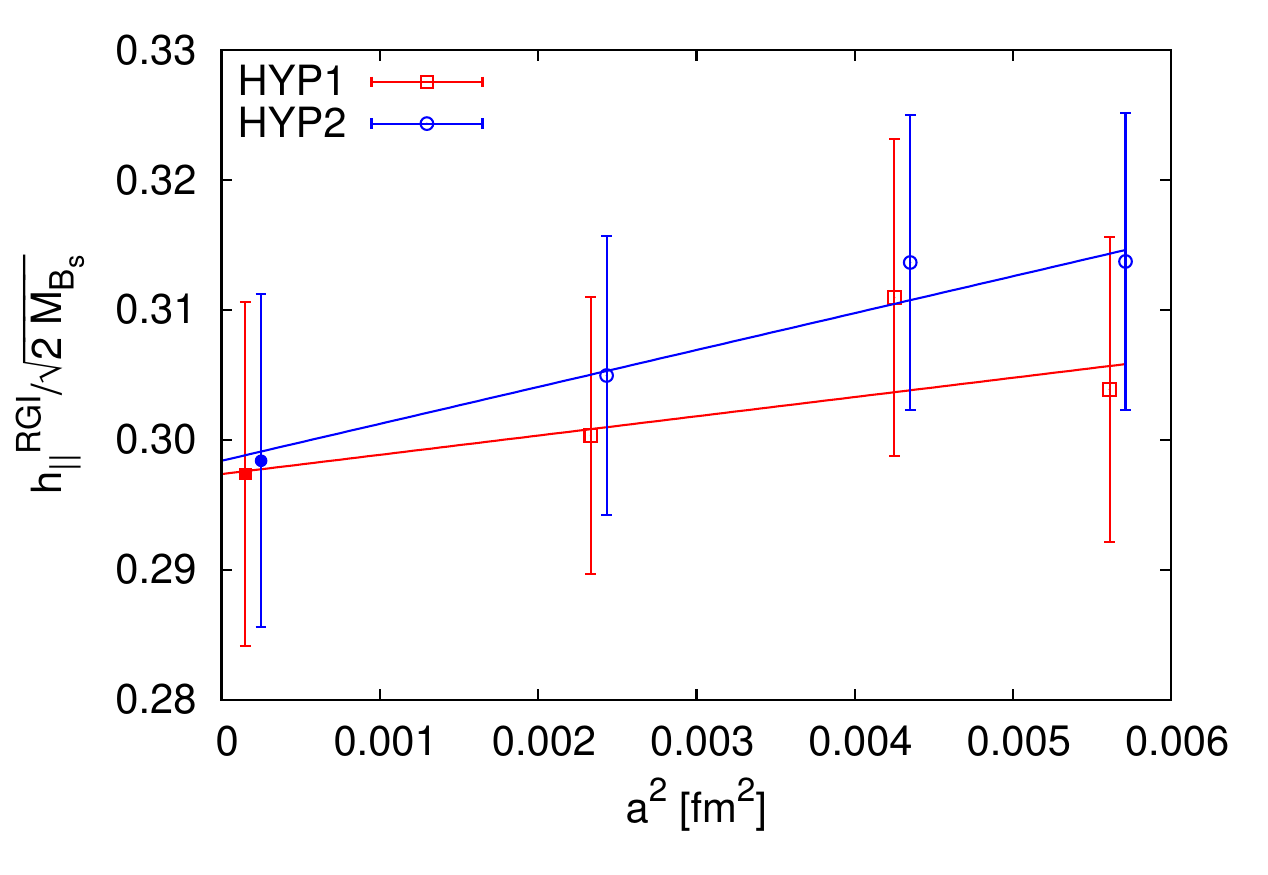} }
\hspace{0cm}\scalebox{0.57}{\includegraphics{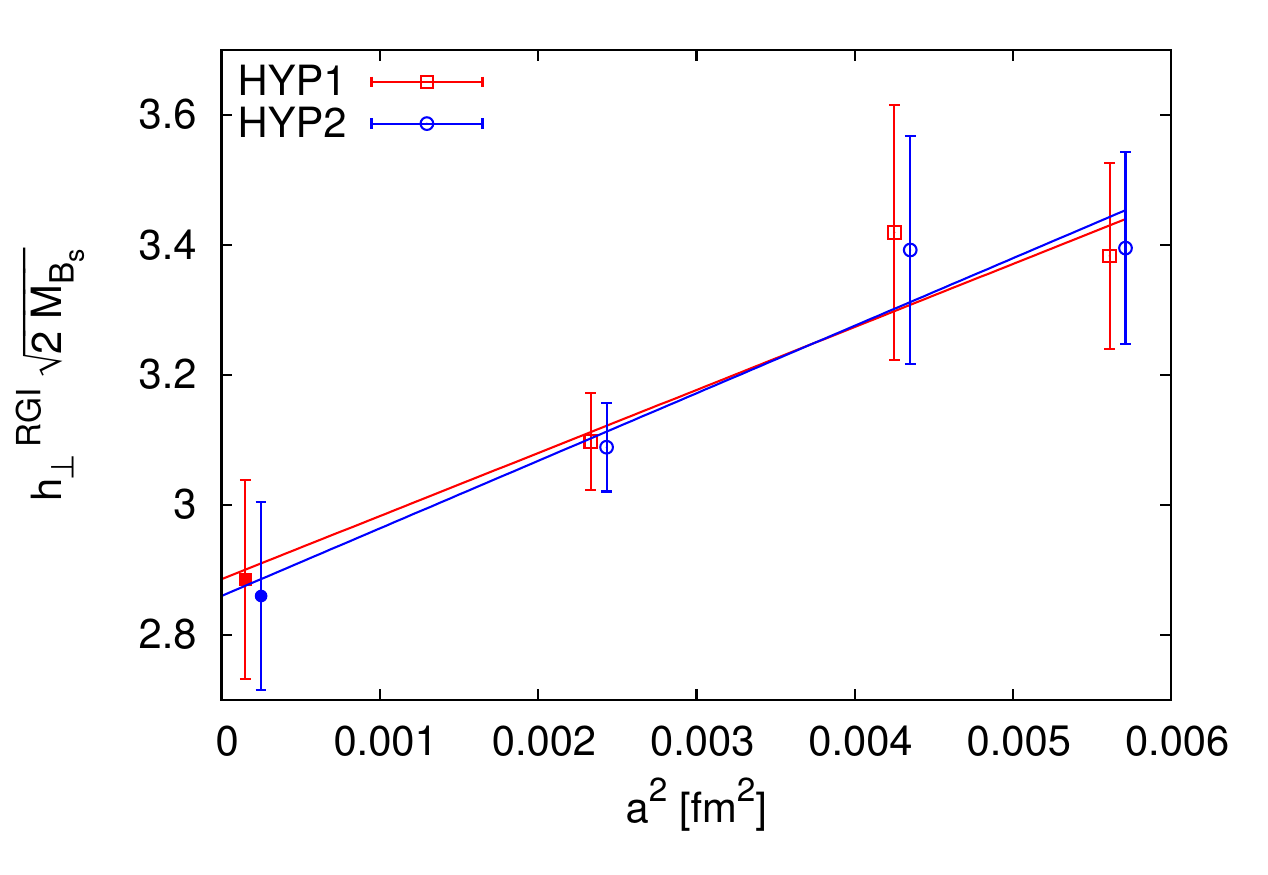} }
\caption{Results for $\fparallel^\RGI$ (left) and $\fperp^\RGI$ (right) 
with 1-loop O$(a)$ improvement coefficients. Data for actions HYP1/2 
are separated slightly in $a^2$ for better visibility. 
The lines show continuum extrapolations linear in $a^2$.
}
\label{f:contlim}
\end{center}
\end{figure}

\fig{contlim} shows the dependence of the results on the lattice spacing
and the continuum extrapolation for the two discretisations (HYP1 and HYP2).
As our best result we estimate the continuum limit by a linear extrapolation 
in $a^2$ of the data with $c_x=c_x^{(1)}g_0^2$, as illustrated in the figure.

For $\fparallel$ a simple constant extrapolation (a weighted average)
yields compatible results, but of course with much smaller error bars. 
Since there is no reason, why $a^2$ effects should be entirely absent,
we use the numbers with the larger error bars. 

It seems not critical that we know the O$(a)$ improvement coefficients 
of the currents only in 1-loop perturbation theory. These coefficients are 
not very relevant at the level of precision of our data.\footnote{The reader
should not be confused by the fact that bare numbers may depend significantly
on $c_x$; discretisation errors have to be assessed after renormalization.
}

Finally we combine the continuum limits of HYP1 and HYP2 in a weighted 
average,
\begin{subequations}
\begin{align}
\fparallel^\RGI  &= 0.976(41)\;\GeV^{1/2}, \label{e:resfpar}  
\\
\fperp^\RGI      &= 0.876(43)\;\GeV^{-1/2} \,.\label{e:resfperp}
\end{align}
\label{e:resfrgi}
\end{subequations}

\subsection{The Form Factors $f_+$ and $f_0$, and a Comparison to Other Results}

We now have different options to estimate the form factor $f_+$. 
Working only at static order, we can use
any quantity $\fplusnum{i}$ with
\bes
  f_+  = \fplusnum{i} \cdot \left[ 1 + \rmO(1/\mbeauty) \right] \;, 
\ees
and we consider (dropping the argument of $C_\mathrm{V_0},C_\mathrm{V_k}$)
\begin{subequations}
\begin{align}
\fplusnum{1}  &= \sqrt{m_\B/2}\,\left( \big(1-\frac{E_\K}{m_\B}\big) \,C_\mathrm{V_k}\,\fperp^\RGI(E_\K) +
  \frac{1}{m_\B}\,C_\mathrm{V_0}\,\fparallel^\RGI(E_\K)\right)\,,
\label{e:fplus1}  
\\
\fplusnum{2}    &= \sqrt{m_\B/2} \,C_\mathrm{V_k}\,\fperp^\RGI(E_\K) 
\,.\label{e:fplus2}
\end{align}
\label{e:resfplus}
\end{subequations}
In $\fplusnum{1}$ all known terms and kinematical factors are taken into 
account, despite the 
fact that the form factors $\fparallel$ and $\fperp$ contain further $1/\mbeauty$ suppressed contributions 
which we do not control, while in $\fplusnum{2}$ we systematically 
drop all $1/\mbeauty$ suppressed contributions. Numerically we have
(combined HYP1/2)
\begin{subequations}
\begin{align}
\fplusnum{1}  &= 1.77(7)(7)\,,
\label{e:fplus1num}  
\\
\fplusnum{2}        &= 1.63(8)(6)
\,,\label{e:fplus2num}
\end{align}
\end{subequations}
where the second errors are the ones from the perturbative uncertainty
in \eqref{e:Cv0} and \eqref{e:Cvk}.

Of course one could also include in \eqref{e:fplus2} the exactly 
known kinematical prefactor of $\fperp$ via
$\fplusnum{3} = (1-{E_\K}/{m_\B})\,\fplusnum{2} = 0.864\times\fplusnum{2}$.
Such $\sim 15\%$ uncertainties/ambiguities will be reduced to $1\%\!-\!2\%$ 
when we include all $1/\mbeauty$ terms.

In order to give a single number, we estimate the O$(1/\mbeauty)$ terms in this way
and quote 
\bes
  f_+(21.22\,\GeV^2) = \fplusnum{2} \pm 0.15 \, \fplusnum{2} = 1.63(8)(6)\pm 0.24\,.
  \label{e:fplusfin}
\ees

Besides $f_+$ it is common in the literature to report results for 
the scalar form factor $f_0$, which is another linear combination
of $\fperp$ and $\fparallel$. To estimate its value, we use
\beq
\fzeronum{1}  = \frac{\sqrt{2/m_\B}}{1-m_\K^2/m_\B^2}\, \left( 
  \big(1-\frac{E_\K}{m_\B}\big) \,C_\mathrm{V_0}\,\fparallel^\RGI(E_\K) +
  \frac{\vecp_\K^2}{m_\B}\,C_\mathrm{V_k}\,\fperp^\RGI(E_\K)\right)\,,
\label{e:fzero1}  
\eeq
where, analogous to $\fplusnum{1}$, all known kinematic O$(1/\mbeauty)$ terms are included.
Our (combined HYP1/2) result is
\beq
\fzeronum{1} = 0.66(3)(1)\,.
\label{e:zero1num}  
\eeq


Our results, \eq{fplusfin} and \eq{zero1num}, compare rather well with 
other values of the form factors in the literature. 
The result of Flynn et al. \cite{Flynn:2015mha}, extracted at
our value of $q^2$, is  $f_+ \approx 1.65 (10)$ and $f_0 \approx 0.62 (5) $, 
while Bouchard et al. \cite{Bouchard:2014ypa} have
$f_+ \approx  1.80 (20) $ and $f_0 \approx 0.66 (5) $.
As should be clear, our estimates have a systematic error of a completely 
different nature. While we focused our effort on taking the continuum
limit of non-perturbatively renormalized matrix elements at a fixed
Kaon momentum, we have so far neglected the dependence on the light-quark mass 
which -- according to \cite{Flynn:2015mha, Bouchard:2014ypa} -- is below our errors.

\section{Conclusion and Outlook}
\label{sec:Conclion_outlook}

For the first time we have been able to perform a study of the
continuum limit of fully non-perturbatively renormalized 
form factors. They are computed at a fixed squared momentum 
transfer $q^2=21.22\,\GeV^2$, and we have concentrated on the 
leading-order form factors in $\minv$. In RGI form these are 
unambiguous. 
Our main result is contained in \fig{contlim}.
It shows that the continuum limit at a Kaon momentum
of $1/2\,\GeV$ is smooth. 
This behavior of the discretisation errors for matrix elements
with a momentum of this size is not obvious a priori and 
linked to our choice 
of actions; a generalization is at most possible at a rather qualitative 
level. With this encouraging result, we may consider also somewhat 
larger momenta in the future.

Our numbers in \eq{fplusfin} and \eq{zero1num} provide a positive
cross-check of {\cite{Flynn:2015mha, Bouchard:2014ypa}. 
Already now, they thus strengthen our confidence in the form factors
extracted on the lattice and summarized in \cite{Aoki:2013ldr},
but they will be of a more direct phenomenological 
interest when the $1/\mbeauty$ terms are included and the errors shrink
accordingly. At that point we also have to carefully consider the 
extrapolation to the physical light quark masses and finally more than one 
value of the Kaon momentum will be of interest.
As a bottom line, the study presented here suggests that all this is possible 
with a precision which is of interest for an extraction of $V_\mathrm{ub}$ from
experimental decay rates.

\begin{acknowledgement}
We thank Alberto Ramos for his contributions in earlier stages of the project.
We had many useful discussions with Michele Della Morte, Piotr Korcyl
and Sara Collins.

We gratefully acknowledge the Gauss Centre for Supercomputing (GCS)
for providing computing time through the John von Neumann Institute for
Computing (NIC) on the GCS share of the supercomputer JUQUEEN at J\"ulich
Supercomputing Centre (JSC). GCS is the alliance of the three national
supercomputing centres HLRS (Universit\"at Stuttgart), JSC (Forschungszentrum
J\"ulich), and LRZ (Bayerische Akademie der Wissenschaften), funded by the
German Federal Ministry of Education and Research (BMBF) and the German State
Ministries for Research of Baden-W\"urttemberg (MWK), Bayern (StMWFK) and
Nordrhein-Westfalen (MIWF).
We acknowledge PRACE for awarding us access to resource JUQUEEN in Germany at J\"ulich and to resource SuperMUC in Germany at M\"unchen.
We also thank the LRZ for a CPU time grant on SuperMUC, project pr85ju, and
DESY for access to the PAX cluster in Zeuthen.

\end{acknowledgement}

\providecommand{\href}[2]{#2}\begingroup\raggedright\endgroup

\end{document}